\documentclass[pra,superscriptaddress,showpacs,twocolumn]{revtex4}

\usepackage{amsfonts}
\usepackage{amsmath}
\usepackage{amssymb}
\usepackage{url}  %% URL formatting
\usepackage{epic} %% enhance the picture environment
\usepackage{bm}   %% provides access to bold math symbols

\begin{document}

\title{Structure of Two-qubit Symmetric Informationally Complete POVMs}
\author{Huangjun Zhu}
\affiliation{Centre for Quantum Technologies, %
National University of Singapore, Singapore 117543, Singapore}
\affiliation{NUS Graduate School for Integrative Sciences and
Engineering, Singapore 117597, Singapore}

\author{Yong Siah Teo}
\affiliation{Centre for Quantum Technologies, %
National University of Singapore, Singapore 117543, Singapore}
\affiliation{NUS Graduate School for Integrative Sciences and
Engineering, Singapore 117597, Singapore}

\author{Berthold-Georg Englert}
\affiliation{Centre for Quantum Technologies, %
National University of Singapore, Singapore 117543, Singapore}
\affiliation{Department of Physics, %
National University of Singapore, Singapore 117542, Singapore}


%03.65.-w quantum mechanics
%03.65.Wj quantum tomography, state reconstruction
%03.67.-a quantum information
%02.10.De algebraic structure

\begin{abstract}
In the four-dimensional Hilbert space, there exist 16
Heisenberg--Weyl (HW) covariant symmetric informationally complete
positive operator valued measures (SIC~POVMs) consisting of 256
fiducial states on a single orbit of the Clifford group. We explore
the structure of these SIC~POVMs by studying the symmetry
transformations within a given SIC~POVM and among different
SIC~POVMs. Furthermore, we find  16 additional SIC~POVMs by a
regrouping of the 256 fiducial states, and show that they are
unitarily equivalent to the original 16 SIC~POVMs by establishing an
explicit unitary transformation. We then reveal the additional
structure of these SIC~POVMs when the four-dimensional Hilbert space
is taken as the tensor product of two qubit Hilbert spaces. In
particular, when either the standard product basis or the Bell basis
are chosen as the defining basis of the HW group, in eight of the 16
HW covariant SIC~POVMs,  all  fiducial states  have the same
concurrence of $\sqrt{2/5}$. These SIC~POVMs are particularly
appealing for an experimental implementation, since all fiducial
states can be connected to each other with just local unitary
transformations. In addition, we introduce a concise representation
of the fiducial states with the aid of a suitable tabular
arrangement of their parameters.
\end{abstract}

\date{\today}
\maketitle

\section{Introduction}
A positive operator valued measure (POVM)  consists of a set of
outcomes represented mathematically as a set of positive operators
that sum up to the identity.  An \emph{informationally complete}
(IC) POVM   allows us to  reconstruct any quantum state from the
probabilities  of outcomes. An IC~POVM contains at least $d^2$
outcomes in a $d$-dimensional Hilbert space. A \emph{minimal} IC
POVM contains exactly $d^2$ outcomes.

A \emph{symmetric informationally complete} (SIC) POVM  \cite{Zau99,
RBSC04, App05,SG09},  which consists of $d^2$ pure subnormalized
projectors with equal pairwise fidelity, stands out as a fiducial
POVM due to its high symmetry and high tomographic efficiency
\cite{Fuc02,RBSC04,App05, Sco06}.  It is generally believed that
SIC~POVMs exist in any Hilbert spaces of finite dimensions since
Zauner's conjecture \cite{Zau99}, although a rigorous mathematical
proof is not known. Up to now, analytical solutions have been found
in dimensions 2, 3 \cite{DGS75}; 4, 5 \cite{Zau99}; 6 \cite{Gra04};
7 \cite{App05}; 8 \cite{Hog98, Gra05}; 9--15 \cite{SG09,
Gra05,Gra06,Gra08a,Gra08b}; 19 \cite{App05}; 24, 35, 48 \cite{SG09}.
Numerical solutions with high precision have also been obtained up
to $d=67$ \cite{RBSC04, SG09}.

The interest in SIC~POVMs extends well beyond their application in
quantum state tomography. The relation between SIC~POVMs and
mutually unbiased bases (MUB) is another focus of  ongoing efforts
\cite{Woo04, ADF07, App08}. In the mathematical community, SIC~POVMs
are  studied under the name of equiangular lines \cite{LS73} and as
minimal 2-designs \cite{RBSC04}. The Lie algebraic significance of
SIC~POVMs was also explored recently \cite{AFF09}.

A \emph{group-covariant} SIC~POVM   can be generated from a single
reference state---\emph{a fiducial state}---with transformations
from a unitary group. Most known SIC~POVMs are covariant with
respect to the \emph{Heisenberg--Weyl (HW) group} or generalized
Pauli group \cite{Zau99, RBSC04, App05}. The structure of the  HW
covariant SIC~POVMs can be studied with the aid  of the normalizer
of the HW group---the \emph{Clifford group} (or the \emph{extended
Clifford group} when antiunitary operations are also included),
which divides the fiducial states and SIC~POVMs into disjoint orbits
\cite{App05, App09}. SIC~POVMs on the same orbit of the extended
Clifford group are unitarily or antiunitarily \emph{equivalent} in
the sense that they can be transformed into each other with unitary
or antiunitary operations \cite{App05}. The equivalence relation
among SIC~POVMs on different orbits is  still an open problem for
arbitrary dimensions. Recently this problem was solved for prime
dimensions \cite{Zhu10}.

In this paper, we focus on HW covariant SIC~POVMs in the
four-dimensional Hilbert space which exhibit  remarkable additional
symmetry beyond what is reflected in the name. According to the
numerical calculations by Renes {\it et al.} \cite{RBSC04} as well
as by Scott and Grassl \cite{SG09}, there exists a single orbit of
256 fiducial states, constituting 16 SIC~POVMs. We shall
characterize these fiducial states and SIC~POVMs by studying the
symmetry transformations within a given SIC~POVM and among different
SIC~POVMs. The symmetry group of each SIC~POVM is shown to be a
subgroup of the Clifford group, thereby extending recent results on
prime dimensions \cite{Zhu10}. Furthermore, we find  16 additional
SIC~POVMs by a regrouping of the 256 fiducial states, and show that
they are unitarily equivalent to the original 16 SIC~POVMs by
establishing an explicit unitary transformation. These additional
SIC~POVMs from a regrouping of fiducial states have also been
noticed by Grassl \cite{Gra08a}.

We then reveal  the additional structure of these SIC~POVMs when the
four-dimensional Hilbert space is taken as the tensor product of two
qubit Hilbert spaces.  A concise representation of the fiducial
states is introduced in terms of the generalized Bloch vectors,
which allows us to explore the intriguing symmetry of the two-qubit
SIC POVMs. In particular, when either the standard product basis or
the Bell basis is chosen as the defining basis of the HW group, in
eight of the 16 HW covariant SIC~POVMs, all the fiducial states have
the same concurrence of $\sqrt{2/5}$; hence these fiducial states
can be turned into each other with just local unitary
transformations. These SIC~POVMs are particularly appealing for an
experimental implementation, because local unitary transformations
are much easier to realize than global ones.

The  paper is organized as follows. In Sec.~\ref{sec:pre}, we set
the stage by recalling basic properties  of SIC~POVMs and Clifford
groups. In Sec.~\ref{sec:structure}, we then study the structure of
SIC~POVMs in the four-dimensional Hilbert space, and construct the
16 additional SIC~POVMs by a regrouping of the fiducial states. In
Sec. \ref{sec:twoqubitSIC}, we deal with the structure of two-qubit
SIC~POVMs. We conclude with a summary.

\section{\label{sec:pre}Setting the stage}
A SIC~POVM \cite{Zau99, RBSC04, App05, SG09},
$\sum_{j=1}^{d^2}\Pi_j=I$ is composed of $d^2$ outcomes that are
subnormalized projectors,
$\Pi_j=|\psi_j\rangle\frac{1}{d}\langle\psi_j|$, such that
\begin{eqnarray}\label{eq:SIC}
|\langle\psi_j|\psi_k\rangle|^2=\frac{1+d\delta_{j k}}{d+1}.
\end{eqnarray}
The  symmetry group $\mathrm{G}_{\mathrm{sym}}$ of a SIC~POVM
consists of all unitary  operations that leave the SIC~POVM
invariant, that is, permute the set of outcomes $\Pi_j$. Likewise,
the extended symmetry group $\mathrm{EG}_{\mathrm{sym}}$ is the
larger group that contains also antiunitary operations. A
group-covariant SIC~POVM is one which can be generated from a
fiducial state with a group of unitary operations, such as the
symmetry group of the SIC~POVM.

Since operators which differ only by overall phase factors implement
the same transformation, it is often more convenient to work with
the projective version of the symmetry group and  extended symmetry
group, which are defined as
$\overline{\mathrm{G}}_{\mathrm{sym}}=\mathrm{G}_{\mathrm{sym}}/I(d)$,
$\overline{\mathrm{EG}}_{\mathrm{sym}}=\mathrm{EG}_{\mathrm{sym}}/I(d)$,
where $I(d)$ is the group consisting of operators which are
proportional to the identity operator $I$. Similarly, throughout the
paper, for any unitary group $\mathrm{G}$,  $\overline{\mathrm{G}}$
is used to denote the group obtained from  $\mathrm{G}$ by
identifying elements which differ only by overall phase factors.

Almost all known SIC~POVMs are covariant with respect to the
Heisenberg-Weyl (HW) group or generalized Pauli group $D$. HW group
is generated by the phase operator $Z$ and the cyclic shift operator
$X$ defined by their action on the kets $|e_r\rangle$ of the
``computational basis":
\begin{eqnarray}
Z|e_r\rangle&=&\omega^r|e_r\rangle, \nonumber\\
X|e_r\rangle&=&\left\{ \begin{array}{cl}
  |e_{r+1}\rangle & r=0,1,\ldots,d-2, \nonumber\\
  |e_0\rangle & r=d-1, \\
\end{array}\right.\\
D_{p_1,p_2}&=&\tau^{p_1 p_2}X^{p_1}Z^{p_2},\label{HW}
\end{eqnarray}
where $\omega=e^{2\pi \mathrm{i}/d}, \tau=-e^{\pi \mathrm{i}/d}$,
$p_1,p_2\in Z_d$, and $Z_d$ is the additive group of integer modulo
$d$. The phase factor of $D_{p_1,p_2}$ has been chosen following
Appleby \cite{App05} to simplify the following discussion. As a
consequence of Eq.~(1), a fiducial ket $|\psi\rangle$ of the HW
group obeys
\begin{eqnarray}
|\langle\psi|D_{p_1,p_2}|\psi\rangle|=\frac{1}{\sqrt{d+1}}\quad
\end{eqnarray}
for all $(p_1,p_2)\neq (0,0)$, which are $d^2-1$ equations.

The Clifford group $\mathrm{C}(d)$ is the normalizer of the HW group
that consists of unitary operators. Likewise, the extended Clifford
group $\mathrm{EC}(d)$ is the larger group that contains also
anti-unitary operators. For any operator $U$ in the extended
Clifford group, $U|\psi\rangle$ is a fiducial ket whenever
$|\psi\rangle$ is one. Fiducial states and SIC~POVMs form disjoint
orbits under the action of the extended Clifford group. SIC~POVMs on
the same orbit of the extended Clifford group are unitarily or
antiunitarily equivalent in the sense that they can be transformed
into each other with unitary or antiunitary operations. The
equivalence problem of SIC~POVMs among different orbits is closely
related to the problem of whether the symmetry group of each HW
covariant SIC~POVM is a subgroup of the Clifford group. The two
problems have been solved for all prime dimensions \cite{Zhu10}, but
remains largely open for non-prime dimensions. For $d=4$, there is
no actual equivalence problem, since there is only one orbit of
SIC~POVMs; nevertheless we shall give an affirmative answer to the
other problem in Sec.~\ref{sec:structure}.

To understand the structure of the Clifford group, we need to
introduce some additional concepts. Define
\begin{eqnarray}
\bar{d}=\left\{\begin{array}{cl}
  d & \text{if $d$ is odd}, \\
  2d & \text{if $d$ is even},\\
\end{array}\right.
\end{eqnarray}
and denote by $\mathrm{SL}(2,Z_{\bar{d}})$ the special linear group
consisting of $2\times 2$ matrices
\begin{eqnarray}\label{eq:F}
F=\left(%
\begin{array}{cc}
  \alpha & \beta\\
  \gamma & \delta \\
\end{array}%
\right)
\end{eqnarray}
with entries in $Z_{\bar{d}}$ and determinant 1  mod $\bar{d}$.
Likewise, $\mathrm{ESL}(2,Z_{\bar{d}})$ is the larger group that
contains also the $2\times 2$ matrices with determinant $-1$ mod
$\bar{d}$.   $\mathrm{SL}(2,Z_{\bar{d}})\ltimes (Z_d)^2$ is the
semidirect product group equipped with the following product rule:
\begin{eqnarray}
(F_1,\chi_1)\circ(F_2,\chi_2)=(F_1F_2,\chi_1+F_1\chi_2),
\end{eqnarray}
where $F_1,F_2\in \mathrm{SL}(2,Z_{\bar{d}})$ and $\chi_1, \chi_2\in
(Z_d)^2$. Similarly, $\mathrm{ESL}(2,Z_{\bar{d}})\ltimes (Z_d)^2$ is
the semidirect product group with the same product rule.

The structure of the Clifford group and the extended Clifford group
can best be understood from the following surjective homomorphism
given by Appleby \cite{App05},
\begin{eqnarray}
f_E&:&  \mathrm{ESL}(2,Z_{\bar{d}})\ltimes (Z_d)^2\rightarrow
\overline{\mathrm{EC}}(d), \nonumber\\
&&UD_{\mathbf{p}}U^\dag=\omega^{\langle
\chi,F\mathbf{p}\rangle}D_{F\mathbf{p}}\nonumber\\
\mbox{for}&& U=f_E(F,\chi), \label{homomorphism1}
\end{eqnarray}
where $\langle \mathbf{p},\mathbf{q}\rangle=p_2q_1-p_1q_2$. When $d$
is odd,  $f_E$ is an isomorphism; when $d$ is even, the kernel
contains the following eight elements:
\begin{eqnarray}
\left(\left(%
\begin{array}{cc}
  1+rd & sd \\
  td & 1+rd \\
\end{array}%
\right),
\left(%
\begin{array}{c}
  sd/2 \\
  td/2 \\
\end{array}%
\right)\right)\quad\mbox{for}\quad r,s,t=0,1.\label{kernel}
\end{eqnarray}
If $\det(F)=1 \mod \bar{d}$ (see  Eq.~(\ref{eq:F}) for the
definition of $F$), and if $\beta$ is invertible in $Z_{\bar{d}}$,
the explicit homomorphism is given by \cite{App05}
\begin{eqnarray}
&(F,\chi)\rightarrow U=D_\chi V_F,&\nonumber\\
&\displaystyle{V_F=\frac{1}{\sqrt{d}}\sum\limits
_{r,s=0}^{d-1}|e_r\rangle\tau^{\beta^{-1}(\alpha s^2-2rs+\delta
r^2)}\langle e_s|.}&\label{eq:Cliffordunitary}
\end{eqnarray}
If $\beta$ is not invertible, there always exists an integer $x$
such that $\delta+x\beta$ is invertible, and $F$ can be written as
the product of two matrices $F=F_1F_2$, where
\begin{eqnarray}
F_1=\left(
      \begin{array}{cc}
        0 & -1 \\
        1 & x \\
      \end{array}
    \right),\quad
F_2=\left(
      \begin{array}{cc}
        \gamma+x\alpha & \delta+x\beta \\
        -\alpha & -\beta \\
      \end{array}
    \right),
\end{eqnarray}
such that $V_{F_1}$ and $V_{F_2}$ can be computed according to
Eq.~(\ref{eq:Cliffordunitary}), then  $V_F=V_{F_1}V_{F_2}$
\cite{App05}.

If $\mathrm{det}(F)=-1$, then $\mathrm{det}(FJ)=1$ and $(FJ,\chi)\in
\mathrm{SL}(2,\bar{d})\ltimes (Z_d)^2$, where
\begin{eqnarray}
J=\left(
    \begin{array}{cc}
      1 & 0 \\
      0 & -1 \\
    \end{array}
  \right).
\end{eqnarray}
Hence the homomorphism images of the elements in
$\mathrm{ESL}(2,\bar{d})\ltimes(Z_d)^2$ can be determined once the
images of the  elements in $\mathrm{SL}(2,p)\ltimes(Z_p)^2$  and
that of $(J,\bm{0})$  are determined respectively, where $\bm{0}$ is
a shorthand for ${0\choose 0}$. The homomorphism image of
$(J,\bm{0})$ is the complex conjugation operator $\hat{J}$
\cite{App05},
\begin{eqnarray}\label{eq:conjugation}
&\hat{J}: \;\sum\limits_{r=0}^{d-1}|e_r\rangle a_r\mapsto\sum\limits_{r=0}^{d-1}|e_r\rangle a_r^*,&%\\
\end{eqnarray}
which is clearly basis-dependent (here defined with respect to the
computational basis) and has no physical meaning. Following Appleby,
$[F,\chi]$ (This is not a commutator!) is used to denote the
homomorphism image of $(F,\chi)$ throughout the paper.

The rest of the paper focuses on the HW covariant SIC~POVMs for
$d=4$ unless  otherwise stated.

\section{\label{sec:structure}   Structure of  SIC~POVMs in the four-dimensional Hilbert space}

For $d=4$, the order of the Clifford group is 768, and that of the
extended Clifford group is 1536. Numerical searches performed by
Renes {\it et al.} \cite{RBSC04} as well as by Scott and Grassl
\cite{SG09} suggest that there is only one orbit of fiducial states
(both under the Clifford group and the extended Clifford group). In
the following discussion, we assume that their numerical searches
are exhaustive.

One of the fiducial states is \cite{App05}
$\rho_\mathrm{f}=|\psi_\mathrm{f}\rangle\langle\psi_\mathrm{f}|$
with
\begin{eqnarray}\label{eq:fiducial}
|\psi_{\mathrm{f}}\rangle&=&(|e_0\rangle, |e_1\rangle, |e_2\rangle,
|e_3\rangle)\frac{1}{2\sqrt{3+G}} \nonumber\\
&&\times\left(
  \begin{array}{c}
    1+e^{-\mathrm{i}\pi/4} \\
     e^{\mathrm{i}\pi/4}+\mathrm{i}G^{-3/2} \\
     1-e^{-\mathrm{i}\pi/4} \\
    e^{\mathrm{i}\pi/4}-\mathrm{i}G^{-3/2} \\
  \end{array}
\right),
\end{eqnarray}
where $G=(\sqrt{5}-1)/2$ is the golden ratio. The stability group
(within the extended Clifford group) of this fiducial state is the
order-6 cyclic group generated by the following antiunitary
operator,
\begin{eqnarray}\label{eq:StaAU}
&&[A_4,\chi_4]=\left[\left(%
\begin{array}{cc}
  -1 & 1 \\
  -1 & 2 \\
\end{array}%
\right) ,\left(%
\begin{array}{c}
  2 \\
  0 \\
\end{array}%
\right) \right]=V\hat{J},
 \end{eqnarray}
 where
\begin{eqnarray}
V\hat{=}\frac{1}{2}\left(
     \begin{array}{cccc}
       1 & e^{\mathrm{i}\pi/4} & -1 &e^{\mathrm{i}\pi/4}\\
       \mathrm{i}& e^{-3\mathrm{i}\pi/4} & \mathrm{i} & e^{\mathrm{i}\pi/4} \\
       1 &  e^{-3\mathrm{i}\pi/4} & -1 &  e^{-3\mathrm{i}\pi/4} \\
       \mathrm{i} &  e^{\mathrm{i}\pi/4} & \mathrm{i} & e^{-3\mathrm{i}\pi/4} \\
     \end{array}
   \right),
\end{eqnarray}
and $\hat{J}$ is the complex conjugation operator defined in
Eq.~(\ref{eq:conjugation}). Within the Clifford group, the stability
group is generated by $[A_4,\chi_4]^2$. Hence there are 256 fiducial
states constituting 16 SIC~POVMs on the orbit \cite{App05,RBSC04}.

\subsection{\label{sec:SICsym} Symmetry transformations within an HW covariant SIC~POVM}
In this section, we focus on the symmetry property of a single HW
covariant SIC~POVM for $d=4$. In particular, we show that the
symmetry group of each HW covariant SIC~POVM is a subgroup of the
Clifford group, and each HW covariant SIC~POVM is covariant with
respect to a unique HW group.

Since all SIC~POVMs are on the same orbit, it is enough to study the
SIC~POVM generated from the fiducial state with the ket in
Eq.~(\ref{eq:fiducial}) under the action of the HW group. To
demonstrate that  the symmetry group $\mathrm{G}_{\mathrm{sym}}$
(extended symmetry group $\mathrm{EG}_{\mathrm{sym}}$) of this
SIC~POVM is a subgroup of the  Clifford group (extended Clifford
group), it is enough to show that the stability group of the
fiducial state $\rho_{\mathrm{f}}$ within the symmetry group is the
same as that within the Clifford group, which is generated by
$[A_4,\chi_4]^2$.

To simplify the notation in the following discussion, we use the
ordered pair $(p_1,p_2)$ to represent the fiducial state with the
ket $D_{p_1,p_2}|\psi_{\mathrm{f}}\rangle$. Under the action of
$[A_4,\chi_4]^2$, the 15 fiducial states other than
$\rho_{\mathrm{f}}\hat{=}(0,0)$ in the SIC~POVM form five orbits:
\begin{eqnarray}
O_1&=&\{(1,0), (0,3), (3,1)\},\nonumber\\
O_2&=&\{(3,3), (3,2), (2,3)\},\nonumber\\
O_3&=&\{(0,1), (1,3), (3,0)\},\nonumber\\
O_4&=&\{(1,2), (2,1), (1,1)\},\nonumber\\
O_5&=&\{(2,0), (0,2), (2,2)\}.
\end{eqnarray}
Now let  $\rho_{j_1}, \rho_{j_2}, \rho_{j_3}$ be any triple of
different fiducial states in the SIC~POVM. Notice that the triple
product traces $\mathrm{tr}(\rho_{j_1}\rho_{j_2}\rho_{j_3})$ are
invariant under unitary transformations. Hence any unitary
transformation in the stability group (within the symmetry group of
the SIC~POVM) of $\rho_{\mathrm{f}}$  must preserve these
invariants. However, at least one of these invariants would be
violated, if there exists any unitary transformation in the
stability group other than that generated by $[A_4,\chi_4]^2$. In
conclusion, the symmetry group of each HW covariant SIC~POVM is a
subgroup of the Clifford group for $d=4$.

From the previous discussions, the order of the symmetry group
$\overline{\mathrm{G}}_{\mathrm{sym}}$ (extended symmetry group
$\overline{\mathrm{EG}}_{\mathrm{sym}}$) of each SIC~POVM is 48
(96), which is much smaller than that of the symmetry group of a
15-dimensional regular simplex. Moreover, it is not always possible
to transform a pair of fiducial states to another pair with either a
unitary or antiunitary operation within the extended symmetry group.
Note that the HW group is a normal Sylow 2-subgroup of
$\overline{\mathrm{G}}_{\mathrm{sym}}$ (for any prime $p$, a Sylow
$p$-subgroup of a finite group is a subgroup of order $p^n$ such
that $p^n$ is the largest power of $p$ that divides the order of the
group \cite{KS04}), hence there is only one Sylow 2-subgroup (here
an order-16 subgroup) in $\overline{\mathrm{G}}_{\mathrm{sym}}$
according to Sylow's theorem. As a result, each HW covariant
SIC~POVM is covariant with respect to a unique HW group for $d=4$.
This observation extends the  previous result on prime dimensions
not equal to three \cite{Zhu10}.

Within each HW covariant SIC~POVM, the triple product traces
$\mathrm{tr}(\rho_{j_1} \rho_{j_2}\rho_{j_3})$ may take on 17
different values (eight pairs of conjugates and one real number). As
we shall see shortly, there exists a continuous  family of
inequivalent triples of normalized states with equal pairwise
fidelity of 1/5. In most cases, it is impossible to extend such a
triple to a full HW covariant SIC~POVM, even in principle, in
contrast with the situation for $d=2$ or $d=3$. This phenomenon is
common in Hilbert spaces of higher dimensions.

In a $d$-dimensional Hilbert space with $d\geq3$,  the three states
corresponding to the three kets, respectively, in the following
equation have equal pairwise fidelity of ${1/(d+1)}$:
\begin{eqnarray}\label{eq:triple}
|\varphi_1\rangle\hat{=}\left(
                 \begin{array}{c}
                   1 \\
                   0 \\
                   0 \\
                   \vdots\\
                   0\\
                          \end{array}
               \right),\;
|\varphi_2\rangle\hat{=}\left(
                 \begin{array}{c}
                   \frac{1}{\sqrt{d+1}} \\
                    \frac{\sqrt{d}}{\sqrt{d+1}} \\
                   0 \\
                   \vdots\\
                   0\\
                            \end{array}
               \right),\;
|\varphi_3\rangle\hat{=}\left(
                 \begin{array}{c}
                    \frac{1}{\sqrt{d+1}} \\
                   u(\theta) e^{i\theta} \\
                   v(\theta) \\
                   \vdots\\
                   0\\
                            \end{array}
               \right),\nonumber\\
\end{eqnarray}
where
\begin{eqnarray}\label{triplefamily2}
u(\theta)&=&\frac{-\cos\theta+\sqrt{(\cos\theta)^2+d}}{\sqrt{d(d+1)}},\nonumber\\
v(\theta)&=&\sqrt{\frac{d^2-d-2(\cos\theta)^2+2\cos\theta\sqrt{(\cos\theta)^2+d}}{d(d+1)}}\nonumber\\
\end{eqnarray}
with $-\pi\leq \theta< \pi$ . Let
\begin{eqnarray}
\phi(\theta)&=&\mathrm{arg}\bigl\{\mathrm{tr}[(|\varphi_1\rangle\langle\varphi_1|)(|\varphi_2\rangle\langle\varphi_2|)(|\varphi_3\rangle\langle\varphi_3|)]\bigr\}\nonumber\\
&=&\mathrm{arg}\Bigl[\frac{1+e^{\mathrm{i}\theta}(-\cos\theta+\sqrt{(\cos\theta)^2+d})}{(d+1)^2}\Bigr].
\end{eqnarray}
There is a one-to-one correspondence between $\theta$ and
$\phi(\theta)$, and  the angle $\phi(\theta)$ of the triple product
may also take on any value between $-\pi$ and $\pi$. Since this
angle is invariant under unitary transformations, any two different
ordered triples of states in the family defined by
Eq.~(\ref{eq:triple}) are inequivalent. On the other hand, any two
ordered triples can be turned into each other by a suitable unitary
transformation  if  the angles of the respective triple product
traces are equal.

For $d= 2$, any triple of states is uniquely specified by the
pairwise fidelity, up to unitary transformations. And the triple can
always be extended to a full SIC~POVM if the pairwise fidelity is
$1/(d+1)$. For $d=3$, there exists a continuous family of
inequivalent SIC~POVMs \cite{Zau99, RBSC04, App05}; with a suitable
choice of SIC~POVMs and fiducial states, the angle of the triple
product trace may take on any value between $-\pi$ and $\pi$
\cite{Zhu10}. Hence any triple of states with equal pairwise
fidelity of $1/(d+1)$ can be extended to a full HW covariant
SIC~POVM.

For $d=4$, as far as the SIC~POVM generated from the fiducial state
in Eq.~(\ref{eq:fiducial}) is concerned, the angle of the triple
product trace of distinct fiducial states may only take on 17
different values. Hence, it is impossible to extend a generic triple
of states with equal pairwise fidelity of $1/(d+1)$  to a full HW
covariant SIC~POVM. The same conclusion also holds for $d>4$ if
there are only a finite number of inequivalent HW covariant
SIC~POVMs, which seems to be the case according to the numerical
searches performed by Scott and Grassl \cite{SG09}.

Similarly, it is reasonable to expect that it is generally
impossible to extend a given set of $k$ states (${k<d^2}$) with
equal pairwise fidelity of $1/(d+1)$ to a full SIC~POVM. This
observation implies that we need some global constraints in addition
to  the local constraint of equal pairwise fidelity to fully
characterize the structure of a SIC~POVM. It also illustrates the
difficulty of constructing a SIC~POVM through adding states one by
one in a successive manner.

\subsection{Reconstruction of the HW group
from a given SIC~POVM}\label{sec:reconHW}

Most known examples of SIC~POVMs are constructed from  fiducial
states under the action of the HW group. In this section we
investigate the inverse problem: reconstruct the HW group from a
given SIC~POVM. This problem is relevant in determining whether a
SIC~POVM constructed with a different method is covariant with
respect to the HW group, and in determining the equivalence relation
among SIC~POVMs, as we shall see in Sec.~\ref{sec:EXSIC}. Our
approach is described for  SIC~POVMs in the four-dimensional Hilbert
space; however, it can be generalized to SIC~POVMs in  Hilbert
spaces of other dimensions  if certain conditions are satisfied.

Let $M$ be the sum of four  different fiducial states in the
SIC~POVM generated from the fiducial state in
Eq.~(\ref{eq:fiducial}),
$M=\rho_{j_1}+\rho_{j_2}+\rho_{j_3}+\rho_{j_4}$. Our reconstruction
scheme is based on the set of eigenvalues and eigenkets of $M$.
Since the SIC~POVM is group-covariant and eigenvalues are invariant
under unitary transformations, we can assume that
$\rho_{j_1}=\rho_\mathrm{f}$ without loss of generality. When the
four fiducial states are related by the transformation $Z$, we have
\begin{eqnarray}\label{eq:reconEig}
M&=&\sum_{j=0}^{3} Z^j\rho_{\mathrm{f}}
Z^{-j}=\sum_{j=0}^{3}|e_j\rangle\lambda_j\langle e_j|,\nonumber\\
\lambda_{0,2}&=&\frac{1}{\sqrt{5}}(2\pm \sqrt{2})G, \quad
\lambda_{1,3}=\frac{2}{\sqrt{5}G}\pm\sqrt{\frac{2}{5G}}.
\end{eqnarray}
If   $\rho_{\mathrm{f}}$ is replaced by another fiducial state, then
the order of the diagonal entries of $M$ may be changed, however,
$\lambda_0$ and $\lambda_2$ never appear in adjacent position in the
diagonal of $M$, neither do $\lambda_1$ and $\lambda_3$. In
addition, since the generator of the stabilizer of
$\rho_{\mathrm{f}}$ in Eq.~(\ref{eq:StaAU}) implements the following
cyclic transformation:
\begin{eqnarray}
Z\rightarrow XZ^2\rightarrow XZ^3\rightarrow X^2Z\rightarrow
X^3\rightarrow XZ\rightarrow Z,
\end{eqnarray}
the set of eigenvalues of $M$ is the same if the four states
$\rho_{j_1},\rho_{j_2},\rho_{j_3},\rho_{j_4}$ are related by
 any  order-4 element in the HW group.
Further calculation shows that the set of eigenvalues of $M$ will be
different if the four states cannot be connected by any order-4
element in the HW group.

We are now ready to reconstruct the HW group from a given SIC~POVM
based on the previous observations:
\begin{enumerate}
  \item Find four different fiducial states such that the set of  eigenvalues of their sum $M$  is the same as that given in
Eq.~(\ref{eq:reconEig}), and calculate the normalized eigenkets
$|e_k^\prime\rangle$  of $M$ corresponding to the eigenvalues
$\lambda_k$ for $k=0,1,2,3$ respectively. Then the operator
$Z^\prime=\sum_{k=0}^3 |e_k^\prime\rangle\mathrm{i}^k\langle
e_k^\prime|$ is a generator of the HW group to be reconstructed.

  \item Under the action of
  $Z^\prime$, the 16 fiducial states form four orbits of equal length. Choose four fiducial states, one  from each orbit, such
  that the four states possess the same property as in the first step,
  then construct another operator $X^\prime$ as in the first step.

  \item The group generated by the two operators $X^\prime, Z^\prime$ is exactly the
  HW group of interest.
\end{enumerate}

In addition to reconstructing the HW group from a given HW covariant
SIC~POVM, the above procedure can also be applied to check the group
covariance of a SIC~POVM constructed with a different method. If we
cannot find the set of fiducial states required in Steps 1 or 2, or
the group thus reconstructed is not unitarily equivalent to the HW
group, or the SIC~POVM is not covariant under the group thus
constructed (if such a SIC~POVM exists), then the SIC~POVM is not HW
covariant.

\subsection{\label{sec:symTran} Symmetry transformations among HW covariant SIC~POVMs}
In this section, we study the symmetry transformations among the 16
HW covariant SIC~POVMs for $d=4$ and reveal the structure underlying
these SIC~POVMs.

To describe the symmetry transformations among the 16 SIC~POVMs, we
first need to label each SIC~POVM with a unique number for later
reference. Define $V_n=[F_n,\bm{0}]$ for $n=1,2,\ldots,16$, where
$F_n$s are given by
\begin{eqnarray}\label{Eq:Fn}
&&\left(
\begin{array}{cc}
 1 & 0 \\
 0 & 1
\end{array}
\right),\quad \left(
\begin{array}{cc}
 0 & 3 \\
 5 & 7
\end{array}
\right),\quad\left(
\begin{array}{cc}
 2 & 1 \\
 1 & 1
\end{array}
\right),\quad \left(
\begin{array}{cc}
 6 & 7 \\
 3 & 5
\end{array}
\right),\nonumber\\
&&\left(
\begin{array}{cc}
 0 & 3 \\
 5 & 5
\end{array}
\right),\quad\left(
\begin{array}{cc}
 0 & 1 \\
 7 & 1
\end{array}
\right),\quad \left(
\begin{array}{cc}
 6 & 7 \\
 7 & 7
\end{array}
\right),\quad\left(
\begin{array}{cc}
 3 & 1 \\
 1 & 6
\end{array}
\right), \nonumber\\
 &&\left(
\begin{array}{cc}
 3 & 1 \\
 2 & 1
\end{array}
\right),\quad\left(
\begin{array}{cc}
 6 & 7 \\
 1 & 4
\end{array}
\right),\quad\left(
\begin{array}{cc}
 0 & 3 \\
 5 & 6
\end{array}
\right),\quad\left(
\begin{array}{cc}
 0 & 1 \\
 7 & 0
\end{array}
\right),\nonumber\\ &&\left(
\begin{array}{cc}
 6 & 7 \\
 5 & 6
\end{array}
\right),\quad\left(
\begin{array}{cc}
 3 & 1 \\
 0 & 3
\end{array}
\right),\quad\left(
\begin{array}{cc}
 0 & 1 \\
 7 & 2
\end{array}
\right),\quad\left(
\begin{array}{cc}
 0 & 3 \\
 5 & 0
\end{array}
\right).
\end{eqnarray}
Let the image of the SIC~POVM containing the fiducial state in
Eq.~(\ref{eq:fiducial}) under the transformation $V_n$ be SIC~POVM
No.~$n$, then this correspondence between the 16 HW covariant SIC
POVMs and the 16 numbers  $n=1,2,\ldots,16$ is one-to-one. Here the
$F_n$s have been chosen in foresight to simplify the following
discussion.
\begin{table}
\caption{\label{tab:hierarchy1} Arrangement of the 16 HW covariant
SIC~POVMs for $d=4$. Each number $n$, with $1\leq n\leq 16$,
represents the HW covariant SIC~POVM obtained by transforming the
SIC~POVM containing the  fiducial state in Eq.~(\ref{eq:fiducial})
with the unitary transformation $[F_n,\bm{0}]$ specified in
Eq.~(\ref{Eq:Fn}). } \vspace{1ex} \centering
\begin{math}
\begin{array}{cccc}
1&2&3&4\\
5&6&7&8\\
9&10&11&12\\
13&14&15&16\\
\end{array}
\end{math}
\end{table}

\begin{figure}
\begin{center}
    \begin{picture}(200,410)%(40,0)
%%%%%%%%%%%%%%%%%%%%%%%%%%%%%%%%%%%%%%%%%%%%%%%%%%%%%%%%%%%%%%%%%%%%%%
%% plot A
\matrixput(0,330)(16,0){4}(0,16){4}{\circle*{2} }
\matrixput(0,332)(16,0){4}(0,32){2}{\vector(0,1){12.5} }
\matrixput(0,344)(16,0){4}(0,32){2}{\vector(0,-1){12.5} } %%
%%%%%%%%%%%%%%%%%%%%%%%%%%%%%%%%%%%%%%%%%%%%%%%%%%%%%%%%%%%%%%%%%%%%%%
%% plot B
\matrixput(74,330)(16,0){4}(0,16){4}{\circle*{2} }
\matrixput(75.5,331.5)(32,0){2}(0,32){2}{\vector(1,1){13} }
\matrixput(88.5,344.5)(32,0){2}(0,32){2}{\vector(-1,-1){13} }
\matrixput(88.5,331.5)(32,0){2}(0,32){2}{\vector(-1,1){13} }
\matrixput(75.5,344.5)(32,0){2}(0,32){2}{\vector(1,-1){13} } %%
%%%%%%%%%%%%%%%%%%%%%%%%%%%%%%%%%%%%%%%%%%%%%%%%%%%%%%%%%%%%%%%%%%%%%%%
%% plot C
\matrixput(148,330)(16,0){4}(0,16){4}{\circle*{2} }
\matrixput(150,330)(32,0){2}(0,16){4}{\vector(1,0){12.5} }
\matrixput(162,330)(32,0){2}(0,16){4}{\vector(-1,0){12.5} } %%
%%%%%%%%%%%%%%%%%%%%%%%%%%%%%%%%%%%%%%%%%%%%%%%%%%%%%%%%%%%%%%%%%%%%%%%
%% plot D
\matrixput(0,220)(25,0){4}(0,25){4}{\circle*{2} }
\matrixput(49,222)(25,0){2}(0,25){4}{\vector(-1,0){23} }
\multiput(26,218)(0,25){4}{\vector(1,0){48} } %%
%%%%%%%%%%%%%%%%%%%%%%%%%%%%%%%%%%%%%%%%%%%%%%%%%%%%%%%%%%%%%%%%%%%%%%%
%% plot E
\matrixput(120,220)(25,0){4}(0,25){4}{\circle*{2} }
\multiput(120,222)(0,50){2}{\vector(0,1){21} }
\multiput(120,243)(0,50){2}{\vector(0,-1){21} }
\multiput(146,221.5)(0,50){2}{\vector(1,1){22} } %%
\multiput(172.5,220.5)(0,50){2}{\vector(1,1){22} } %%
\multiput(146,243.5)(0,50){2}{\vector(1,-1){22} } %%
\multiput(172.5,244.5)(0,50){2}{\vector(1,-1){22} } %%
\multiput(193.5,221.5)(0,50){2}{\vector(-2,1){47} } %%
\multiput(193.5,243.5)(0,50){2}{\vector(-2,-1){47} } %%
%%%%%%%%%%%%%%%%%%%%%%%%%%%%%%%%%%%%%%%%%%%%%%%%%%%%%%%%%%%%%%%%%%%%%%%%%
%% plot F
\matrixput(0,110)(25,0){4}(0,25){4}{\circle*{2} } %%
\put(3,185){\vector(1,-1){47}}%%
\put(0,157){\vector(1,-1){47}}%%
\multiput(24.2, 158.4)(0,25){2}{\vector(-1,-2){23}}%%
\multiput(50.8, 158.4)(0,25){2}{\vector(1,-2){23}}%%
\put(75, 156.5){\vector(-1,-1){47}}%%
\put(72, 185){\vector(-1,-1){47}}%%
%%%%%%%%%%%%%%%%%%%%%%%%%%%%%%%%%%%%%%%%%
\put(1.6, 135.8){\vector(2,1){47}}%%
\put(25, 138){\vector(-1,1){22}}%%
\put(50, 138){\vector(1,1){22}}%%
\put(73.4, 135.8){\vector(-2,1){47}}%%
%%%%%%%%%%%%%%%%%%%%%%%%%%%%%%%%%%%%%%%%
\put(3, 111){\vector(2,3){47}}%%
\put(24.5, 111.5){\vector(-1,3){24}}%%
\put(50, 111.5){\vector(1,3){24.2}}%%
\put(72.5, 111){\vector(-2,3){47.5}}%%
%%%%%%%%%%%%%%%%%%%%%%%%%%%%%%%%%%%%%%%%%%%%%%%%%%%%%%%%%%%%%%%%%%%%%%%%%%%
%% plot G
\matrixput(120,110)(25,0){4}(0,25){4}{\circle*{2} } %%
\multiput(116, 112)(75,0){2}{\vector(0,1){72}}%%
\multiput(118, 137)(75,0){2}{\vector(0,1){22}}%%
\multiput(122, 158)(75,0){2}{\vector(0,-1){47}}%%
\multiput(124, 183)(75,0){2}{\vector(0,-1){47}}%%
\put(145,158.4){\vector(1,-2){23.3}}%%
\put(146.6,183.4){\vector(1,-2){23.3}}%%
\put(170,158.4){\vector(-1,-2){23.3}}%%
\put(168.4,183.4){\vector(-1,-2){23.3}}%%

\put(145.6,111.8){\vector(1,3){23.3}}%%
\put(169.4,111.8){\vector(-1,3){23.3}}%%
\put(146.5,136.5){\vector(1,1){23.3}}%%
\put(168.5,136.5){\vector(-1,1){23.3}}%%
%%%%%%%%%%%%%%%%%%%%%%%%%%%%%%%%%%%%%%%%%%%%%%%%%%%%%%%%%%%%%%%%%%%%%%%%%%%%%%%%
%% plot H
\matrixput(0,0)(25,0){4}(0,25){4}{\circle*{2} } %%
\put(1.5,50.5){\vector(3,1){72}} %%
\put(73.5,74.5){\vector(-3,-1){72}}%%
 \put(1.5,74.5){\vector(3,-1){72}}%%
\put(73.5,50.5){\vector(-3,1){72}}%%
\put(26.5,51.5){\vector(1,1){22}} %%
\put(48.5,73.5){\vector(-1,-1){22}}%%
 \put(26.5,73.5){\vector(1,-1){22}}%%
\put(48.5,51.5){\vector(-1,1){22}}%%
\multiput(2,-2)(0,25){2}{\vector(1,0){72}} %%
\multiput(73,-2)(0,25){2}{\vector(-1,0){72}}%%
\multiput(27,2)(0,25){2}{\vector(1,0){22}}%%
\multiput(48,2)(0,25){2}{\vector(-1,0){22}}%%
%%%%%%%%%%%%%%%%%%%%%%%%%%%%%%%%%%%%%%%%%%%%%%%%%%%%%%%%%%%%%%%%%%%%%%%%%%%%%%%%%
%% plot I
\matrixput(120,0)(25,0){4}(0,25){4}{\circle*{2} } %%
\multiput(120,2)(25,0){4}{\vector(0,1){21.5}}%%
\multiput(120,23)(25,0){4}{\vector(0,-1){21.5}}%%
%%%%%%%%%%%%%%%%%%%%%%%%%%%%%%%%%%%%%%%%%%%%%%%%%%%%%%%%%%%%%%%%%%%%%%%%%%%%%%%%
%% plot labels
          \put(0,85){\mbox{H}}
          \put(120,85){\mbox{I}}
           \put(0,195){\mbox{F}}
          \put(120,195){\mbox{G}}
          \put(0,305){\mbox{D}}
          \put(120,305){\mbox{E}}
           \put(0,385){\mbox{A}}
          \put(75,385){\mbox{B}}
          \put(148,385){\mbox{C}}
\end{picture}
\end{center}
        \caption{\label{fig:SymmetryT} Illustration of the symmetry transformations among the 16 HW covariant
        SIC~POVMs induced by elements in the group $\mathrm{EG}_{\mathrm{SYM}}=\overline{\mathrm{EC}}(d)/\overline{D}$ (see Sec.~\ref{sec:symTran}).
        Here, every dot represents a SIC~POVM arranged as in Table~\ref{tab:hierarchy1}, and every arrow starts from a
        SIC~POVM before the symmetry transformation and ends at another SIC~POVM after the
        symmetry transformation. Only one element in each conjugacy class of $\mathrm{G}_{\mathrm{SYM}}$ is chosen as a representative,
         the transformations induced by other elements within the same conjugacy class can be obtained by permuting the columns.
         In the case of order-4 elements, only two out of the four conjugacy classes are chosen; the elements in the other two conjugacy classes
         are the inverses of the elements in these two conjugacy classes respectively, so their transformations can be obtained
         by reversing the arrows.  Plot A:  order-2  element
        in the center of $\mathrm{G}_{\mathrm{SYM}}$;
        plots B and C: another two order-2
        elements from the other two  conjugacy classes  respectively; plots D and  E: an order-3 element and an order-6
        element respectively; plots F and  G: two
        order-4 elements from two different conjugacy classes
        respectively; plot H: the complex
        conjugation operation; plot I: the complex conjugation operation followed by an appropriate order-2 element in $\mathrm{G}_{\mathrm{SYM}}$.
        }
\end{figure}
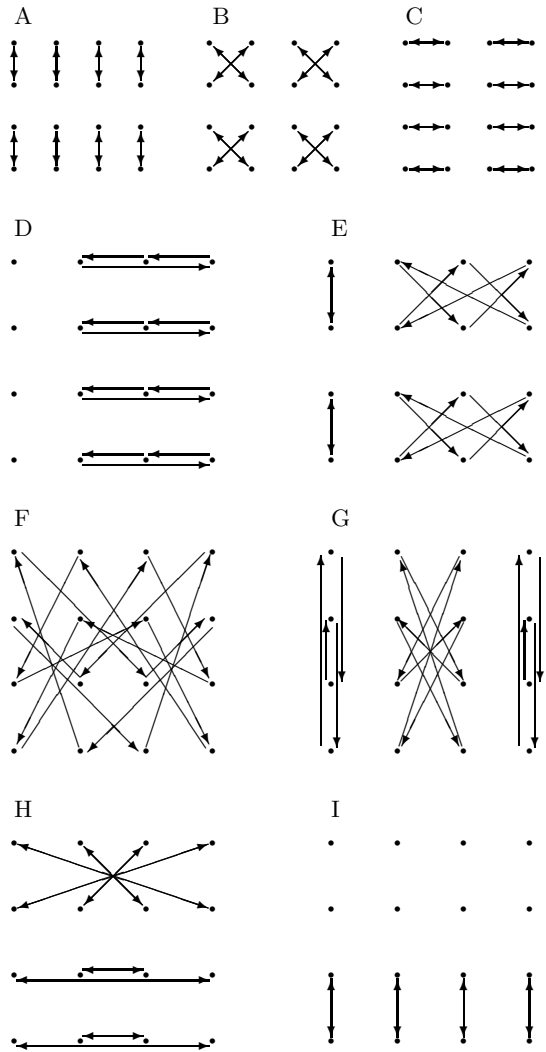

Since we are now only  concerned with the transformations among
different SIC~POVMs, the groups
$\mathrm{G}_{\mathrm{SYM}}=\mathrm{\overline{C}}(4)/\overline{D}$
and
$\mathrm{EG}_{\mathrm{SYM}}=\mathrm{\overline{EC}}(4)/\overline{D}$
properly describe the symmetry operations we consider. As an
abstract group, $\mathrm{G}_{\mathrm{SYM}}$ is isomorphic to the
special linear group $\mathrm{SL}(2,4)$ defined in
Sec.~\ref{sec:pre}; likewise $\mathrm{EG}_{\mathrm{SYM}}$ is
isomorphic to the extended special linear group $\mathrm{ESL}(2,4)$.
Coincidently, the order of $\mathrm{G}_{\mathrm{SYM}}$ is the same
as the order of the symmetry group
$\overline{\mathrm{G}}_{\mathrm{sym}}$ of a single SIC~POVM, that is
48; however, the two groups are not isomorphic. The group
$\mathrm{G}_{\mathrm{SYM}}$ consists of the identity, seven order-2
elements, eight order-3 elements, 24 order-4 elements and eight
order-6 elements. There are three conjugacy classes for order-2
elements, with one, three and three elements respectively. There are
four conjugacy classes for order-4 elements, each with six elements;
elements in two of the classes are the inverses of the elements in
the other two classes respectively. There is only one conjugacy
class for either order-3 elements or order-6 elements. The center of
$\mathrm{G}_{\mathrm{SYM}}$ is generated by the order-2 element
which has only one conjugate.

If  the 16 HW covariant SIC~POVMs are arranged in a $4\times 4$
square as in Table~\ref{tab:hierarchy1}, then the effect of the
symmetry transformations of the group $\mathrm{G}_{\mathrm{SYM}}$
can be delineated in a pictorial way as shown in
Fig.~\ref{fig:SymmetryT}. The effect of only one group element in
each conjugacy class is shown; the effect of other group elements
within the same conjugacy class can be obtained simply by permuting
the columns representing the SIC~POVMs.

According to Fig.~\ref{fig:SymmetryT}, the symmetry transformations
among the 16 SIC~POVMs can be decomposed into row transformations
and column transformations. In addition to the identity, all order-3
elements and one class of order-2 elements (see plots D and C in
Fig.~\ref{fig:SymmetryT}) transform the SIC~POVMs within each row,
and with the same effect in every row. They constitute an order-12
normal subgroup of $\mathrm{G}_{\mathrm{SYM}}$, which can also be
identified as the alternating group of the four columns. The
quotient group of $\mathrm{G}_{\mathrm{SYM}}$ with respect to this
group of row transformations can then be identified with an order-4
cyclic subgroup (generated by the cyclic permutation of the four
rows $1\rightarrow3\rightarrow2\rightarrow4\rightarrow1$, see plots
F and G in Fig.~\ref{fig:SymmetryT}) of the symmetry group of the
four rows. Similarly, the quotient group of
$\mathrm{EG}_{\mathrm{SYM}}$ can be identified with an order-8
subgroup of the symmetry group of the four rows.

\subsection{\label{sec:EXSIC}  Additional SIC~POVMs from regrouping of the fiducial states}
In this section, we show that there are 16 additional SIC~POVMs from
a regrouping of the 256 fiducial states. These additional SIC~POVMs
are quite peculiar in that they are not constructed from fiducial
states under the action of the HW group. Nevertheless, they are
unitarily equivalent to the original SIC~POVMs as we shall see
shortly. These additional SIC~POVMs from regrouping fiducial states
have been noticed by Grassl, who also showed that they are
 covariant with respect to the HW group but in a different basis
\cite{Gra08a}.

The construction of these additional SIC~POVMs is best illustrated
if the 16 original HW covariant SIC~POVMs are arranged in a
$4\times4$ square as in Table~\ref{tab:hierarchy1}. Under the action
of the abelian subgroup $H=\{I, X^2, Z^2, X^2Z^2\}$ of the HW group,
the 16 fiducial states in each SIC~POVM form four orbits of equal
size. Given four fiducial states in a SIC~POVM connected by $H$,
then in each of the other three SIC~POVMs in the same row, there
exist exactly four fiducial states which are also connected by $H$,
such that the pairwise fidelity between these four fiducial states
and the given four fiducial states is $\frac{1}{5}$---the value
required to form a SIC~POVM for dimension four. It turns out that
the 16 states thus chosen also form a SIC~POVM. In this way,  four
additional SIC~POVMs can be constructed by the regrouping of the
fiducial states in the four original SIC~POVMs in each row, that is,
16 additional SIC~POVMs in total. Moreover, inspection of the
pairwise fidelity among all 256 fiducial states shows that there are
no other SIC~POVMs that can be constructed by regrouping these
fiducial states.

Surprisingly, these additional SIC~POVMs are unitarily equivalent to
the original ones, despite the different method  of construction.
According to the procedure described in Sec.~\ref{sec:reconHW}, we
can reconstruct the HW group $D^\prime$ for these additional
SIC~POVMs which are generated by the following two operators:
\begin{eqnarray}
X^\prime&=&\left[\left(
          \begin{array}{cc}
            3 & 0 \\
            2 & 3 \\
          \end{array}
        \right),\left(
                  \begin{array}{c}
                    0 \\
                    1 \\
                  \end{array}
                \right)\right]\hat{=}\left(%
\begin{array}{cccc}
 1 & 0 & 0 & 0 \\
 0 & 0 & 0 & 1 \\
 0 & 0 & -1 & 0 \\
 0 & -1 & 0 & 0
\end{array}
\right)
,\nonumber\\
Z^\prime &=&\left[\left(
          \begin{array}{cc}
            3 & 2 \\
            0 & 3 \\
          \end{array}
        \right),\left(
                  \begin{array}{c}
                    3 \\
                    0 \\
                  \end{array}
                \right)\right]\nonumber\\
&\hat{=}&\frac{1}{2} \left(
\begin{array}{cccc}
 0 & 1+\mathrm{i} & 0 & -1+\mathrm{i} \\
 1+\mathrm{i} & 0 & -1+\mathrm{i} & 0 \\
 0 & -1+\mathrm{i} & 0 & 1+\mathrm{i} \\
 -1+\mathrm{i} & 0 & 1+\mathrm{i} & 0
\end{array}
\right).
\end{eqnarray}
Note that $D^\prime$ is also a subgroup of the Clifford group of
$D$. Now it is straightforward to verify that the unitary operator
\begin{eqnarray}
&&U\hat{=}\frac{1}{2}\left(
\begin{array}{cccc}
 -\mathrm{i} & -1 & -\mathrm{i} & -1 \\[0.5ex]
 1 & -\mathrm{i} & -1 & \mathrm{i} \\[0.5ex]
 -\mathrm{i} & 1 & -\mathrm{i} & 1 \\[0.5ex]
 1 & \mathrm{i} & -1 & -\mathrm{i}
\end{array}
\right)
\end{eqnarray}
transforms the  standard HW group $D$ into the HW group $D^\prime$
for these additional SIC~POVMs, that is $D^\prime= UDU^\dagger$.
Meanwhile, $U$ is also the unitary transformation between the
original SIC~POVMs and these additional SIC~POVMs.   The fiducial
state $\rho_{\mathrm{f}}$ defined in Eq.~(\ref{eq:fiducial}) remains
invariant under this transformation due to our specific choice of
$U$.

Further analysis shows that there are 32 subgroups  of the Clifford
group which are unitarily equivalent to the HW group $D$ (including
$D$ itself), out of which only $D$ and $D^\prime$ are normal. The
group generated by the Clifford group and $U$ is the normalizer
(within the unitary group) of the Clifford group, of which the
Clifford group is a subgroup with index 2. This group exhausts all
unitary symmetry operations among the 256 fiducial states.

For $d=3$, there exists a continuous family of orbits of SIC POVMs,
and there are 72 fiducial states constituting eight SIC~POVMs on
each generic orbit \cite{Zau99, RBSC04, App05}. 24 additional
SIC~POVMs can be obtained from a suitable regrouping of the 72
fiducial states. However, these additional SIC~POVMs are not
equivalent to the original ones \cite{Zhu10}. For other dimensions,
as far as the SIC~POVMs found by Scott and Grassl \cite{SG09} are
concerned, only for the orbits 8b and 12b (according to the labeling
scheme of Scott and Grassl), additional SIC~POVMs can be obtained by
a suitable regrouping of the fiducial states \cite{ZE}. We are still
trying to understand why these dimensions and orbits of SIC~POVMs
are special in this aspect.

\section{\label{sec:twoqubitSIC}  Two-qubit SIC~POVMs}
In this section we study the additional structure of SIC~POVMs when
the four-dimensional Hilbert space is perceived as a tensor product
of two qubit Hilbert spaces. The appearance of these additional
properties are generally basis dependent, because it matters how the
four-dimensional Hilbert space is tensor-factored into two
two-dimensional spaces. We shall focus on the product basis and the
Bell basis in the following discussion, since the additional
structure is most appealing in these two specific bases.

Before discussing those properties related to the specific bases, we
first mention a result which is basis independent. The average
purity  of the single qubit reduced states of states in any
two-qubit SIC~POVM is $4/5$, that is, the average tangle or squared
concurrence of states in the two-qubit SIC~POVM is $2/5$. More
generally, in a bipartite Hilbert space of subsystem dimensions
$d_1$ and $d_2$ respectively and total dimension $d=d_1d_2$, the
average purity of the reduced states of either party of states in
any SIC~POVM (if such a SIC~POVM exists)  is
$(d_1+d_2)/(d_1d_2+1)$---this value is equal to the average over all
pure states in the bipartite Hilbert space with respect to the Haar
measure \cite{ZS01}.

\subsection{\label{sec:Prod}Two-qubit SIC~POVMs in the product basis}
For a single qubit, any state can be expressed in terms of the
identity operator $I$ and the three Pauli operators $\sigma_j$ for
$j=x, y, z$; the coefficients of expansion define the Bloch vector.
In the case of two qubits, any state $\rho$ can be expressed in
terms of the tensor products of the identity and the Pauli operators
of each qubit respectively:
\begin{eqnarray}
\rho&=&\frac{1}{4}\Big(I\otimes I+\sum_{j=x,y,z} r_j I\otimes
\sigma_j+\sum_{j=x,y,z} s_j
\sigma_j\otimes I\nonumber\\
&&+\sum_{j,k=x,y,z}C_{jk}\sigma_j\otimes\sigma_k\Big).
\end{eqnarray}
Let
\begin{eqnarray}
v&=&\big(r_x,r_y,r_z,s_x,s_y,s_z,C_{xx},c_{xy},C_{xz},C_{yx},C_{yy},C_{yz},\nonumber\\
&&C_{zx},c_{zy},C_{zz}\big)^T;
\end{eqnarray}
in analogy to the case of a single qubit, the vector $v$ will be
referred to as the generalized Bloch vector (GBV) of $\rho$.
Although quite common, this terminology is slightly abusive and
somewhat misleading. The $s$ column and the three columns of $C$ in
Table~\ref{tab:GBV} transform like three-dimensional column vectors
when the first qubit is rotated by  local unitary transformations;
likewise, the $r$ row and the three rows of $C$ are row vectors for
local unitary transformations of the second qubit. In short, the two
single-qubit Bloch vectors are \emph{vectors}, and the two-qubit
``double vector" $C$ is a \emph{dyadic}.

\begin{table}
\centering
 \caption{\label{tab:GBV} Arrangement of the components of the generalized
Bloch vector of each fiducial state.} \vspace{1ex}
\begin{math}
\begin{array}{c|ccc}
\hline\hline
   & r_x & r_y & r_z \\
  \hline
  s_x & C_{xx} & C_{xy} & C_{xz} \\
  s_y & C_{yx} & C_{yy} & C_{yz} \\
  s_z & C_{zx} & C_{zy} & C_{zz} \\
  \hline\hline
\end{array}
\end{math}
\end{table}

The structure of the  GBVs of the fiducial states are best
illustrated if the components are arranged as in
Table~\ref{tab:GBV}. When the standard product basis is chosen as
the defining basis of the HW group, that is,
$|e_0\rangle=|00\rangle,|e_1\rangle=|01\rangle,|e_2\rangle=|10\rangle,|e_3\rangle=|11\rangle$,
 the 256 fiducial states
divide into two classes, according to the structure of their GBVs.
The first class consists of the 128 fiducial states in the first
eight HW covariant SIC~POVMs, and the second class of the  128
states in the last eight SIC~POVMs (according to the labeling scheme
described in Sec.~\ref{sec:symTran}). The structure of the GBV of
each fiducial state in the first class is shown in the top tabular
of Table~\ref{tab:structurep1}, where
\begin{eqnarray}\label{eq:parameter1}
&a,b,\alpha_1,\alpha_2,\alpha_3,\beta_1,\beta_2,\beta_3=\pm1,&\nonumber\\
&\displaystyle{B=\frac{1}{\sqrt{5}},\quad
 A_{\pm}=\frac{\sqrt{1\pm\sqrt{G}}}{\sqrt{5}}.}&
\end{eqnarray}
The eight sign factors
$a,b,\alpha_1,\alpha_2,\alpha_3,\beta_1,\beta_2,\beta_3$ obey the
constraint
\begin{equation}
ab\alpha_1\alpha_2\alpha_3\beta_1\beta_2\beta_3=1.\label{SignConstraintp1}
\end{equation}
There are seven free sign factors, giving a total of  128
combinations of values, and specifying exactly 128  fiducial states
in the first class. In addition, each SIC~POVM in the first class is
specified by the following three sign functions, each  taking a
constant value for the fiducial states in a given SIC~POVM:
\begin{eqnarray}
h_1=b\alpha_2\alpha_3\beta_3, \quad h_2=\alpha_1\alpha_2\alpha_3,
\quad h_3=ab\alpha_1.\label{eq:sfp1}
\end{eqnarray}

Each combination of the eight sign factors which does not satisfy
Eq.~(\ref{SignConstraintp1})  specifies a Hermitian operator $Q$
which is not positive semidefinite. Nevertheless, $Q$ can be written
as the partial transpose (with respect to the computational basis)
of a fiducial state, and satisfies the following 15 equations as
each fiducial state does:
\begin{eqnarray}
\mathrm{tr}(QD_{p_1,p_2}QD_{p_1,p_2}^\dagger)=\frac{1}{5}
\end{eqnarray}
for all $(p_1,p_2)\neq(0,0)$. These equations mean that  the 16
operators generated from $Q$ under the action of the HW group also
form a 15-dimensional regular simplex in the Hilbert space of
Hermitian operators.

\begin{table}
\centering \caption{\label{tab:structurep1} The structure of the
generalized Bloch vector of each fiducial state in the first class
(top) and that in the second class (bottom) when the standard
product basis is chosen as the defining basis
 of the HW group. }
\begin{ruledtabular}
\begin{tabular}{l|ccc}
    & $\beta_1A_b$ & $\beta_2 A_{-b}$ & $\beta_3$ B\\
   \hline
    $\alpha_1 B$ & $\alpha_1\beta_1A_{-b}$ & $\alpha_1\beta_2A_{b}$ & $\alpha_1\beta_3 B$\\
  $\alpha_2 A_a$  & $\sqrt{2}a\alpha_2\beta_1A_a\delta_{a,b}$ & $\sqrt{2}a\alpha_2\beta_2A_a\delta_{a,-b}$ & $\alpha_2\beta_3A_{-a}$\\
  $\alpha_3 A_{-a}$  & $-\sqrt{2}a\alpha_3\beta_1A_{-a}\delta_{-a,b}$ & $-\sqrt{2}a\alpha_3\beta_2A_{-a}\delta_{a,b}$ & $\alpha_3\beta_3A_{a}$\\
\end{tabular}
\end{ruledtabular}
\vspace{2ex}
\begin{ruledtabular}
\begin{tabular}{l|ccc}
   & $\beta_1A_a$ & $\beta_2 A_{a}$ & $\beta_3 B$ \\
   \hline
    $\alpha_1 B$      & $\alpha_1\beta_1A_{-a}$              & $\alpha_1\beta_2A_{-a}$ & $\alpha_1\beta_3 B$ \\
  $\alpha_2 A_a$    & $a^{(1-b)/2}\alpha_2\beta_1 G_{-b}$  & $a^{(1+b)/2}\alpha_2\beta_2G_{b}$  & $\alpha_2\beta_3A_{-a}$ \\
  $\alpha_3 A_{a}\hphantom{{}_-}$  & $a^{(1+b)/2}\alpha_3\beta_1G_b$    & $a^{(1-b)/2}\alpha_3\beta_2G_{-b}$ & $\alpha_3\beta_3A_{-a}$\\
\end{tabular}
\end{ruledtabular}
\end{table}

The structure of the GBV of each fiducial state in the second class
is shown in the bottom tabular of Table~\ref{tab:structurep1}, where
\begin{eqnarray}\label{eq:parameter2}
 &\displaystyle{G_{\pm}=\frac{\sqrt{1\pm G}}{\sqrt{5}}}&
\end{eqnarray}
and $A_{\pm}, B$ are defined in Eq.~(\ref{eq:parameter1}). There is
also one constraint among the eight sign factors, namely
\begin{equation}
b\alpha_1\alpha_2\alpha_3\beta_1\beta_2\beta_3=1.\label{SignConstraintp2}
\end{equation}
Each SIC~POVM in the second class is also specified by  three sign
functions,
\begin{eqnarray}
h_1=ab\alpha_1\beta_3, \quad h_2=-\alpha_1\alpha_2\alpha_3,\quad
h_3=b\alpha_1. \label{eq:sfp2}
\end{eqnarray}

When the SIC~POVMs are arranged as in Table~\ref{tab:hierarchy1} and
Eq.~(\ref{Eq:Fn}), the sign function $h_1$ is a constant in each
row, while the sign functions $h_2, h_3$ are constants in each
column, see Table~\ref{tab:signfunction}. This is one of the reasons
why the numbering in Table~\ref{tab:hierarchy1} was done that way.

\begin{table}
\caption{\label{tab:signfunction} The values of the three sign
functions $h_1,h_2,h_3$ (defined in Eqs.~(\ref{eq:sfp1}) and
(\ref{eq:sfp2})) for each HW covariant SIC~POVM labeled according to
Sec.~\ref{sec:symTran}. }
 \centering
\begin{ruledtabular}
\begin{tabular}{l|cccc}
&$h_2=1$&$h_2=1$&\;\;\;$h_2=-1$&$h_2=-1$\\
&\;\;\;$h_3=-1$&$h_3=1$&$h_3=1$&$h_3=-1$\\
\hline
$h_1=-1$&1&2&3&4\\
$h_1=1$&5&6&7&8\\
$h_1=1$&9&10&11&12\\
$h_1=-1$&13&14&15&16\\
\end{tabular}
\end{ruledtabular}
\end{table}

Since the standard product basis is chosen as the defining basis of
the HW group,  $Z$ and $X^2$ are both local unitary operators. The
16 fiducial states in each SIC~POVM  divide into two sets of equal
size, such that the eight fiducial states in each set have the same
concurrence.
 For each SIC~POVM in the second class, eight
fiducial states have concurrence of $\sqrt{(2+2\sqrt{G})/5}$, and
the other eight have concurrence of $\sqrt{(2-2\sqrt{G})/5}$. What
is peculiar for each SIC~POVM in the first class is that  all 16
fiducial states have the same concurrence of $\sqrt{2/5}$ (tangle of
$2/5$). One could say that these symmetric IC POVMs are not just
symmetric, they are supersymmetric. This supersymmetry is
remarkable, indeed.

Fiducial states in the first class can be turned into each other
with just local unitary transformations. This property is
particularly appealing for an experimental implementation of these
POVMs, because local unitary transformations are much easier to
realize than global ones. On the other hand, the eight SIC~POVMs in
the first class can be transformed into each other with local
Clifford unitary transformations, and so can the eight SIC~POVMs in
the second class.

Since the average tangle of fiducial states in any two-qubit
SIC~POVM is $2/5$, and the concurrence and entanglement of formation
are both concave functions of the tangle, the average concurrence or
entanglement of formation of fiducial states in a SIC~POVM is
maximized when the tangle (concurrence) of each fiducial state is
the same, as for each SIC~POVM in the first class.

Although all fiducial states in each SIC~POVM in the first class
have the same concurrence, nevertheless, it is impossible to connect
all fiducial states  with only local unitary transformations from
the symmetry group $\overline{\mathrm{G}}_{\mathrm{sym}}$ of the
SIC~POVM. The same conclusion also holds for any other basis.
Suppose otherwise, to connect all fiducial states in a SIC~POVM, the
order of the local unitary transformation group is necessarily a
multiple of 16; on the other hand, the order must  divide  the order
of $\overline{\mathrm{G}}_{\mathrm{sym}}$, which is 48. It follows
that the local unitary transformation group must have order either
16 or 48, and thus contains the HW group as a subgroup, since the HW
group is the only order-16 subgroup in
$\overline{\mathrm{G}}_{\mathrm{sym}}$ according to
Sec.~\ref{sec:SICsym}. However, the HW group cannot be a local
unitary group, hence a contradiction would arise.

Furthermore, in each SIC~POVM, exactly two fiducial states have the
same single-qubit reduced states for the first qubit, and the same
is true for the second qubit. The end points of the Bloch vectors of
the eight distinct single-qubit reduced states for each qubit form a
quite regular pattern, especially for the second qubit and for each
SIC~POVM in the first class, where they form a cube.

In bipartite Hilbert spaces, SIC POVMs such that all fiducial states
have the same Schmidt coefficients are quite rare. In
eight-dimensional Hilbert space, there is a SIC~POVM which is
covariant with respect to an alternative version of the HW
group---the three fold tensor product of the Pauli group
\cite{Hog98, Gra05}. Since all fiducial states are connected to each
other by a local unitary group, they have the same Schmidt
coefficients according to any bipartition of the three parties. As
far as the SIC~POVMs found by Scott and Grassl \cite{SG09} are
concerned, which are covariant with respect to the HW group defined
in Eq.~(\ref{HW}), such SIC POVMs only exist on the orbits 4a, 6a,
12b, 28c (according to the labeling scheme of Scott and Grassl).
Interestingly, in all these examples, $d_2=2$ ($d_1=d/d_2=2, 3, 6,
14$); hence concurrence is well defined. According to the discussion
at the beginning of this section, the  purity of the reduced density
matrix of each fiducial state is $(d_1+2)/(2d_1+1)$, thus the
concurrence of each fiducial state is $\sqrt{2(d_1-1)/(2d_1+1)}$.
When $d_1=2$, this is exactly the concurrence of each fiducial state
in the first class of the two-qubit SIC~POVM.

\subsection{\label{sec:Bell}  Two-qubit SIC~POVMs in the Bell basis}
Now consider the Bell basis as the defining basis of the HW group,
that is,
\begin{eqnarray}
|e_0\rangle&=&\frac{1}{\sqrt{2}}(|00\rangle+|11\rangle),\nonumber\\
|e_1\rangle&=&\frac{1}{\sqrt{2}}(|00\rangle-|11\rangle),\nonumber\\
|e_2\rangle&=&\frac{1}{\sqrt{2}}(|01\rangle+|10\rangle),\nonumber\\
|e_3\rangle&=&\frac{1}{\sqrt{2}}(|01\rangle-|10\rangle).
\end{eqnarray}

The structure of the GBV of each fiducial state in the first class
(according to the classification scheme in Sec.~\ref{sec:Prod}) is
shown in the top tabular of Table~\ref{tab:structureB1}, where
$A_{\pm}, B$ are defined in Eq.~(\ref{eq:parameter1}). As in the
case of the product basis, here
$a,b,\alpha_1,\alpha_2,\alpha_3,\beta_1,\beta_2,\beta_3$ may only
take on values $\pm 1$,  and satisfy one constraint,
\begin{eqnarray}\label{SignConstraintB1}
ab\alpha_1\alpha_2\alpha_3\beta_1\beta_2\beta_3=1.
\end{eqnarray}
 In addition, each SIC~POVM is specified by  three sign
 functions,
\begin{eqnarray}
h_1&=&-b\alpha_1\beta_1\beta_2\beta_3,\; h_2=-\beta_1\beta_2\beta_3,
\; h_3=a b\beta_1.
\end{eqnarray}

\begin{table}
\caption{\label{tab:structureB1} \small The structure of the
generalized Bloch vector of each fiducial state in the first class
(top) and that in the second class (bottom) when the Bell basis is
chosen as the defining basis
 of the HW group. } \centering
\begin{ruledtabular}
\begin{tabular}{l|ccc}
   & $\beta_1 B$ & $\sqrt{2}\beta_2A_a \delta_{a,b}$&  $\sqrt{2}\beta_3A_{-a}\delta_{-a,b}$ \\
   \hline
    $\alpha_1 B$& $\alpha_1\beta_1 B$         & $\sqrt{2}\alpha_1\beta_2A_{-a}\delta_{a,b}$  & $\sqrt{2}\alpha_1\beta_3A_{a}\delta_{-a,b}$ \\
  $\alpha_2 A_b \hphantom{{}_-}$& $\alpha_2\beta_1 A_{-b}$   & $b\alpha_2\beta_2 A_a$                       & $b\alpha_2\beta_3 A_{-a}$ \\
  $\alpha_3 A_{b}$ & $\alpha_3\beta_1 A_{-b}$ & $a\alpha_3\beta_2 A_{a}$                     & $-a\alpha_3\beta_3 A_{-a}$ \\
\end{tabular}
\end{ruledtabular}
\vspace{2ex}
\begin{ruledtabular}
\begin{tabular}{l|ccc}
   & $\beta_1 B$ & $\beta_2 G_{-b}$&  $\beta_3 G_{b}$ \\
   \hline
  $\alpha_1 B$       & $\alpha_1\beta_1 B$      & $-b \alpha_1\beta_2 G_{-b}$             & $b\alpha_1\beta_3 G_{b}$ \\
  $\alpha_2 A_{-a}$  & $\alpha_2\beta_1 A_{a}$  & $(-a)^{(1-b)/2} \alpha_2\beta_2 A_{-a}$ & $(-a)^{(1+b)/2}\alpha_2\beta_3 A_{-a}$ \\
  $\alpha_3 A_{a}$   & $\alpha_3\beta_1 A_{-a}$ & $a^{(1-b)/2}\alpha_3\beta_2 A_{a}$     & $a^{(1+b)/2}\alpha_3\beta_3 A_{a}$ \\
\end{tabular}
\end{ruledtabular}
\end{table}

The structure of the GBV of each fiducial state in the second class
is shown in the bottom tabular of Table~\ref{tab:structureB1}, where
$G_{\pm}$ is defined in Eq.~(\ref{eq:parameter2}). Here the sign
factors $a,b,\alpha_1,\alpha_2,\alpha_3,\beta_1,\beta_2,\beta_3$
obey the constraint
\begin{eqnarray}
-ab\alpha_1\alpha_2\alpha_3\beta_1\beta_2\beta_3=1,
\end{eqnarray}
and each SIC~POVM is specified by  three sign functions,
\begin{eqnarray}
h_1=a b\alpha_1,\quad  h_2=-a\beta_1\beta_2\beta_3,\quad
h_3=b\beta_1.
\end{eqnarray}

The values of the three sign functions for each group covariant
SIC~POVM are the same as that in the case of the product basis, see
Table~\ref{tab:signfunction}. In contrast, now fiducial states in
the second class rather than in the first class have the same
concurrence of $\sqrt{2/5}$, while each fiducial state in the first
class may have concurrence of either $\sqrt{(2+2\sqrt{G})/5}$ or
$\sqrt{(2-2\sqrt{G})/5}$.

\section{\label{sec:conclusion} Summary}

We have studied the structure of HW covariant SIC~POVMs in the
four-dimensional Hilbert space, in particular, the symmetry
transformations within one SIC~POVM and among different SIC~POVMs.
The symmetry group of each HW covariant SIC~POVM is shown to be a
subgroup of the Clifford group, extending the previous results on
prime dimensions \cite{Zhu10}. Moreover, we showed that there are 16
additional SIC~POVMs by a suitable regrouping of the 256 fiducial
states, and demonstrated their equivalence with the original 16
SIC~POVMs by  establishing an explicit unitary transformation from
the original SIC~POVMs.

We then revealed the rich  structure of these HW covariant SIC~POVMs
 when the four-dimensional Hilbert space is taken as the
tensor product of two qubit Hilbert spaces. The introduction of
generalized Bloch vectors allowed us to represent the fiducial
states and SIC~POVMs in a very concise way, and to explore their
structure in a systematic manner. In both the product basis and the
Bell basis,  eight of the 16 SIC~POVMs consist of  fiducial states
with the same concurrence of $\sqrt{2/5}$. They are thus not just
symmetric IC~POVMs, but supersymmetric IC~POVMs.

\acknowledgments We are grateful for valuable discussions with
Markus Grassl. This work is supported by the National Research
Foundation and the Ministry of Education, Singapore.

\end{document}